# Non-Gaussian CMBR angular power spectra

J.C.R. Magueijo

Mullard Radio Astronomy Observatory, Cavendish Laboratory
Madingley Road
Cambridge, CB3 0HE, UK
and
Department of Applied Mathematics and Theoretical Physics
University of Cambridge
Cambridge CB3 9EW, UK.

## Abstract

In this paper we show how the prediction of CMBR angular power spectra $C_l$ in non-Gaussian theories is affected by a cosmic covariance problem, that is $(C_l, C_{l'})$ correlations impart features on any observed $C_l$ spectrum which are absent from the average $C^l$ spectrum. Therefore the average spectrum is rendered a bad observational prediction, and two new prediction strategies, better adjusted to these theories, are proposed. In one we search for hidden random indices conditional to which the theory is released from the correlations. Contact with experiment can then be made in the form of the conditional power spectra plus the random index distribution. In another approach we apply to the problem a principal component analysis. We discuss the effect of correlations on the predictivity of non-Gaussian theories. We finish by showing how correlations may be crucial in delineating the borderline between predictions made by non-Gaussian and Gaussian theories. In fact, in some particular theories, correlations may act as powerful non-Gaussianity indicators.

# 1 Introduction

The CMBR temperature fluctuations are often assumed to constitute a 2-D Gaussian random field. If this were the case it is known that the angular power spectrum $C^l$ would fully specify the fluctuations' statistics (see [1] for a review). While Gaussianity is probably a good working hypothesis in the context of inflationary scenarios, it is also accepted that in one way or another the fluctuations predicted in topological defect scenarios are non-Gaussian ([2, 3, 4]). The issue of non-Gaussianity has so far been approached in the form of Gaussianity tests. Topological tests [5], peaks' statistics [1], the 3-point correlation function [6], and skewness and kurtosis tests [7, 8] have been proposed. If these tests showed the fluctuations to be non-Gaussianity then one would have to do more than to measure the $C^l$. A whole set of invariants, components of the $n$-point correlation function ($n > 2$), would then be required in order to fully specify the fluctuations ([9]). Something less obvious is that non-Gaussian statistics would also affect the connection between theoretical and experimental $C^l$, a relationship behind any data-analysis strategy. In this paper we show how this may be the case, taking as an example a texture low-$l$ CMBR model [4] known to display strong non-Gaussian behaviour.

Throughout this paper we will use the notation $C_l$ for the angular power spectrum of realizations, and $C^l$ for its ensemble average. Whereas $C_l$ is a random variable, $C^l$ is a number. Since either $C_l$ or $C^l$ are essentially sets of components of the 2-point correlation function, they cannot by themselves reflect non-Gaussianity. However, the $C_l$ variances involve the $4^{th}$ moments of the $a^l_m$ distribution, dependent on the statistics. Moreover the $C_l$ are necessarily dependent random variables in non-Gaussian theories ([9]). As a result, not only will the cosmic variance in the $C_l$ be affected by non-Gaussianity (usually in the form of a non-Gaussian variance excess), but also **the cosmic variance problem becomes a cosmic covariance problem.** Cosmic covariance can make the $C_l$ vs $C^l$ comparison troublesome. An example will be given in Section 2.1 showing how $C_l$ correlations may impart features on any observed $C_l$ spectra which average out to zero in the $C^l$. Whenever this happens the average $C^l$ spectrum is a bad prediction for the observed CMBR sky. Conversely, the observed $C_l$ also becomes a bad estimator for the $C^l$, a fact already hinted at by the abnormally large cosmic variance in the $C_l$ found in some texture models known to be very non-Gaussian [4].

In this paper we approach the cosmic covariance problem with strategies to do away with the $C_l$ correlations. In Section 2.1 we show how the cosmic variance excess, seemingly connected with correlations, can often be swept under one single variable (the random index, say, $y_1$). We therefore conditionalize the $C_l$ spectrum to this variable, and find that in all the conditionalized sub-ensembles the cosmic variance is highly reduced and the correlations disappear. An analogy with a random tilt Gaussian theory proves to describe realistically what is going on. Hence, instead of using $C^l$ as a prediction for the $C_l$, we advocate the use of the conditionalized spectra $C^l(|y_1)$ together with the random index distribution $F(y_1)$. In Sec. 2.2, on the other hand, we apply to the problem a principal component analysis [10]. We define a new set of $\{\tilde{C}^l\}$, rotated from the original ones, which diagonalize the covariance matrix. In terms of these, realizations $\hat{C}_l$ and averages $\tilde{C}^l$ relate in the usual way. We believe that both methods constitute new, more sensible strategies for connecting theory and experiment in non-Gaussian models. In Section 3 we digress on the implications of what we have said on non-Gaussian cosmic variance to the concept of predictivity. Finally, in Section 4, we point out the key role played by $C_l$ correlations when confronting non-Gaussian theories among themselves and with Gaussian theories. We stress the importance of always considering the $\{C_l\}$ as a whole, and of comparing predictions made by different theories in terms of joint $C_l$ probabilities rather than marginal distributions. We devise an approximation scheme with which to compute the cosmic confusion and preference contours in $\{C_l\}$ space in the presence of correlations. We give examples of strong non-Gaussianity indicators, in the form of pockets in $\{C_l\}$ space, where non-Gaussian theories have high probability over their competing Gaussian counterparts. We close the paper with an outlook of possible applications of what we have said in a more general setting. In particular we suggest that cosmic covariance might dramatically undermine traditional methods for predicting the defect Doppler peak structure.

## 2  Predicting $C_l$ observations in non-Gaussian theories

Topological defect observational predictions are often cast in a language borrowed from inflation. CMBR defect simulations typically output a number of skies to which a spherical harmonics decomposition is applied so as to obtain

$$C_l = \sum_m |a_m^l|^2.$$

Averaging over independent skies one then obtains the angular power spectrum $C^l = <C_l>$. Some simulations (eg. [2]) have qualitatively shown an unusually high cosmic variance $\sigma^2(C_l)$. A quantitative formula for the excess variance (relative to Gaussian theories) was also provided by an analytical model for textures [4]. In Fig. 1 we have plotted the $C^l$ spectrum for a texture model according to [4]. Superposed are the $\pm\sigma/2$ cosmic variance error bars for a Gaussian theory with the same power spectrum and for the texture theory. Notice how large the excess of cosmic variance is for low $l$. What is the origin of this excess variance? Does it imply that the theory is less predictive? Or have we simply applied to the theory an inadequate analysis procedure?

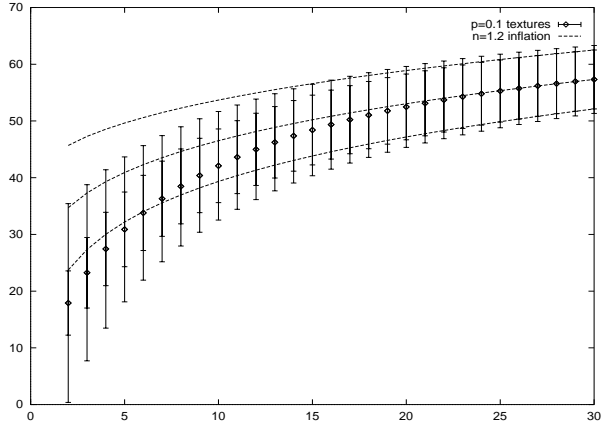

Figure 1: The $C^l$ spectrum for texture Gaussian spots with $p = 0.1$ and $n = 1$. The inner (outer) error bars are Gaussian (full) cosmic variance error bars. We have superposed on it the fitting $n^i = 1.2$ tilted Gaussian theory and associated error bars.

## 2.1 The random index strategy

Consider an hypothetical theory in which the ensemble of all Universes can be split into sub-ensembles which are Gaussian theories. However, let the average $C^l$ spectrum in each sub-ensemble be a random variable from the point of view of the overall ensemble. A concrete realization of this idea is a tilted spectrum Gaussian theory in which the spectral index $n^i$ is a random variable. Then $< C_l(n^i) >_{n^i} = C^l(n^i)$ and $\sigma^2_{n^i}(C_l(n^i)) = C^{l2}(n^i)\frac{2}{2l+1}$, where the subscript $n^i$ means that the averages are taken within a sub-ensemble of constant $n^i$. If $f(n^i)$ is the spectral index distribution one has for the whole ensemble

$$C^l = \int dn^i f(n^i) < C_l(n^i) >_{n^i} = \int dn^i C^l(n^i) = < C^l(n^i) > \quad (1)$$

but now the cosmic variance is

$$\sigma^2(C_l) = \int dn^i f(n^i)\left(< C_l^2(n^i) >_{n^i} - C^{l2}\right) = C^{l2}\frac{2}{2l+1} + \frac{2l+3}{2l+1}\sigma^2(C^l(n^i)) \quad (2)$$

in rough analogy with the texture cosmic variance formula. In such a scenario the $C^l$ spectrum is an unlikely observation. What everyone sees is one of the possible $C^l(n^i)$ spectra, with Gaussian fluctuations around it. Trying to compare the observations with the unphysical $C^l$ is responsible for the cosmic variance excess. It should be obvious that from the point of view of the whole ensemble the $C_l$ cannot be independent, as their individual values give away to some extent the sub-ensemble they were drawn from. One can check that indeed, although $< C_l(n^i)C_{l'}(n^i) >_{n^i} = C^l(n^i)C^{l'}(n^i)$, one has

$$\text{cov}(C_l, C_{l'}) = \int dn^i f(n^i)(C^l(n^i)C^{l'}(n^i) - C^l C^{l'}) \,. \quad (3)$$

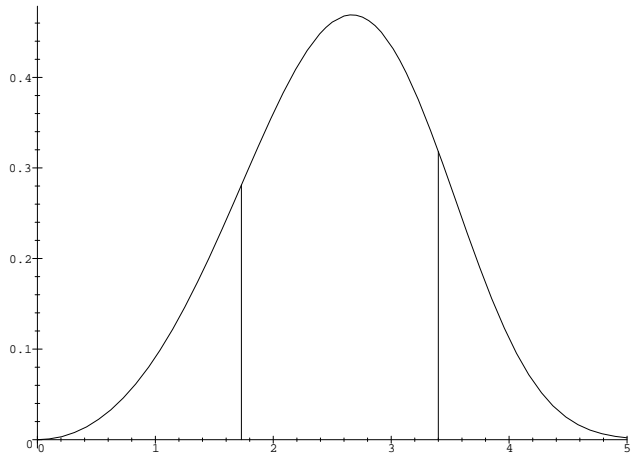

Figure 2: The function $P(y_1)$ for $n = 1$. The bars represent the 68% confidence level interval $(1.7, 3.4)$. The peak is at 2.66.

The suggestion is that $C_l$ excess variance, $C_l$ correlations, and the existence of a random spectral index are related incidences. A random tilt theory is of course less predictive than a fixed tilt theory. Nevertheless, the extra variance is essentially $\sigma^2(n^i)$, not the sum of all the excess variances in the $C_l$, as one might have naively thought. To make the point clear, consider the extreme example of a theory with a deterministic tilt, that is, a theory in which each observer sees a spectrum of the form

$$C_l = \frac{n^i l + c}{l(l+1)} . \qquad (4)$$

Now let $n^i$ be a random variable. A simple minded calculation of the cosmic variance leads to

$$\sigma^2(C_l) = \frac{\sigma^2(n^i)}{(l+1)^2} \qquad (5)$$

which is clearly meaningless, as no randomness is seen by any observer. The uncertainty in the predictions of such a theory is just $\sigma^2(n^i)$.

We will now show that texture theories behave somewhere in between the two examples given above, once one realizes that the sky position of the last texture, $y_1$, acts as a random index. The idea is to conditionalize the $C_l$ to $y_1$ and see what the statistics are in the conditionalized sub-ensembles. Following the notation and results of [4] it is easy to prove that the average $C^l$ spectrum

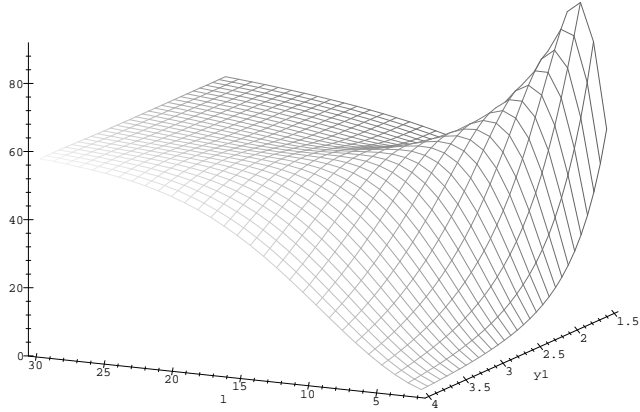

Figure 3: The spectrum $C^l(|y_1)$ for $y_1 \in (1.5, 4)$ for $(n = 1, p_s = 0.1)$ Gaussian spots.

conditional to $y_1$ is

$$C^l(|y_1) = C_1^l(y_1) + \overline{C}_1^l(y_1) = \frac{<a^2>}{4\pi}\left(W^{ls2}(y_1) + \int_{y_1}^{y_{ls}} dy\, N(y) W^{ls2}(y)\right). \quad (6)$$

This is the sum of the last texture brightness plus another term describing the average conditional contribution from all the other textures. The $y_1$ distribution function is

$$P_1(y_1) = N(y_1) e^{-M(y_1)} \quad (7)$$

with $M(y_1) = \int_0^{y_1} dy\, N(y)$. It can be checked that indeed

$$C^l = <C^l(|y_1)> = \int_0^{y_{ls}} dy_1\, P(y_1) C^l(|y_1) = \frac{<a^2>}{4\pi} \int_0^{y_{ls}} dy\, N(y) W^{ls2}(y) \quad (8)$$

in agreement with [4]. We have plotted $P_1(y_1)$ in Fig. 2 for a $n = 1$ texture model. We have also plotted the 68% confidence level interval. Formula (6) suggests that regardless of $y_1$, each observer sees a near white noise regime ($C^l \approx const$) for $l$ up to $l_c$ : $\frac{4\pi}{l_c(l_c+1)} \approx \Omega(y_1)$. The spectrum then crosses over to scale-invariance (or slightly tilted). The cross-over $l_c$ is a random variable, so the white noise feature is absent from the averaged $C^l$, albeit present in any $C^l(|y_1)$. The $C^l(|y_1)$ spectra have been plotted in Fig. 3 for $(n = 1, p_s = 0.1)$ Gaussian spots and should be compared with $C^l$ in Fig. 1. We see that, although $C^l = <C^l(|y_1)>$, the most likely observed spectrum is shaped rather differently from the average $C^l$ spectrum, as in fact are any of the observed spectra. A spectrum like the average $C^l$ is a very improbable observation, a fact which is behind the large cosmic variances in the unconditionalized $C_l$.

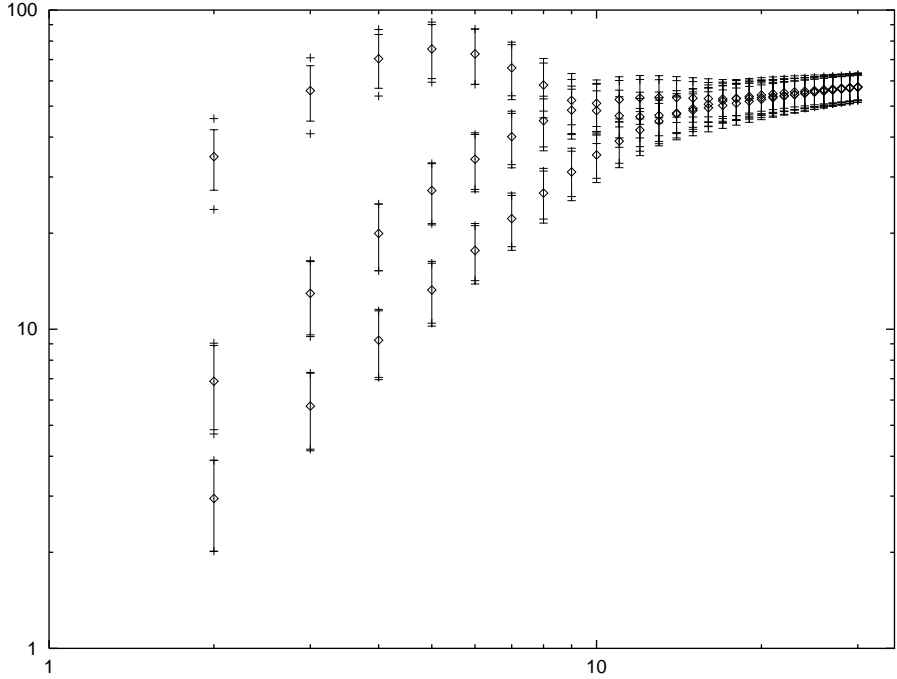

Figure 4: Log-log plot of the $C_l(|y_1)$ spectrum for $y_1$ at the peak (middle spectrum) and borders of 68% confidence level interval (top and bottom). The residual cosmic variance is plotted as error bars, and the points superposed on them correspond to the cosmic variance of a Gaussian theory with a $C^l(|y_1)$ spectrum.

The residual cosmic variance within each $y_1$ ensemble can be computed by introducing the appropriate modifications in the calculation performed in Sec.3.3 in [4] and is

$$\sigma^2_{TX}(C_l(|y_1)) = \overline{C}_1^{l2}(|y_1)\left(\frac{2}{2l+1} + \left(1 + \frac{\sigma^2(a^2)}{<a^2>^2}\right)\frac{\int_{y_1}^{y_{ls}} dy\, N(y)W^{ls4}(y)}{(\int_{y_1}^{y_{ls}} dy\, N(y)W^{ls2}(y))^2}\right)$$
$$+ \frac{4}{2l+1}C_1^l(y_1)\overline{C}_1^l(y_1)\ . \tag{9}$$

In Fig. 4 we have plotted the $C_l(|y_1)$ spectrum with $\pm\sigma/2$ residual cosmic variance error bars for $y_1$ at peak value and borders of 68% confidence level interval. Superposed on these error bars are the corresponding error bars for a Gaussian theory with a $C^l(|y_1)$ spectrum. We see that the residual variance is either comparable to the Gaussian cosmic variance or is smaller. Hence $y_1$ behaves like the random spectral index of the examples above, and the conditionalized theory contains a residual variance somewhere in between the Gaussian and the deterministic examples given. In the next Section we will also show that the conditional correlations are negligible, reinforcing the analogy with a random tilt theory. Overall, it seems wise to take $C^l(|y_1)$ combined with $P(y_1)$, and not $C^l$, as the real prediction for a texture theory.

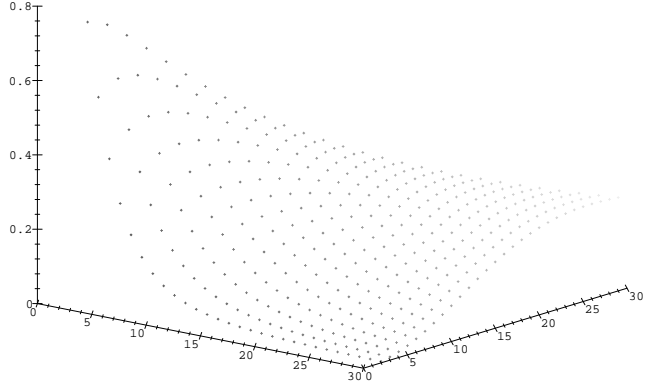

Figure 5: The Pearson's coefficient $cor(C_l, C_{l'})$ for texture Gaussian spots with $p = 0.1$ and $n = 1$.

The example given with textures is probably quite general. Non-Gaussianity in defect theories arises whenever a particular CMBR feature is due to a single defect. Conditionalizing the fluctuations to that defect's variables can significantly reduce the variance. Some guesswork is always required, but a good recipe is to look at the conditionalized cosmic covariance matrix. If there are no residual correlations left then there are no more hidden random indices. All our predictions are then properly parameterized and their variance has been minimized.

We should say that it is not a tragedy that the predicted spectra depend on a parameter which is only statistically predicted by the theory. After all the spectral index $n^i$, related to the shape of the inflationary potential, is an entirely free parameter in inflationary scenarios (within a range dependent on the particular model). The important thing is to realize that the unpredictivity introduced by random indices is the random index variance, not the sum of the variance surplus in all the unconditionalized $C_l$.

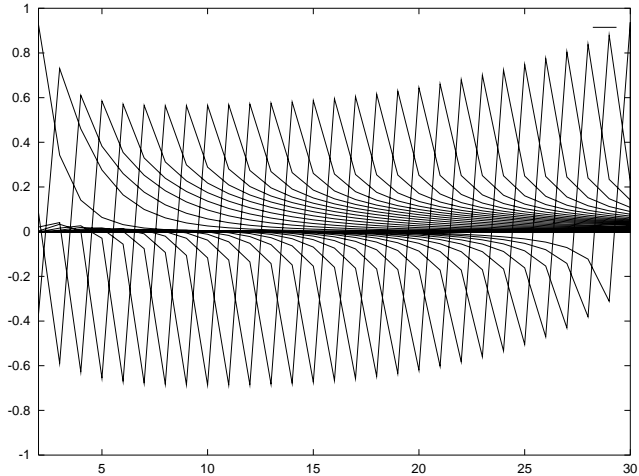

Figure 6: Eigenvector components for texture Gaussian spots with $p = 0.1$ and $n = 1$. The peaks from left to right correspond to $\tilde{C}_l$ in decreasing order of cosmic variance.

## 2.2 $C_l$ covariance matrix and principal components

An alternative method to get rid of the correlations is to apply to the problem a principal component analysis [10]. The idea is to rotate the $\{C_l\}$ frame into a new set $\{\tilde{C}_l\}$ (the principal components) which diagonalizes the covariance matrix. As the $\tilde{C}_l$ are uncorrelated, no features may arise in the observed $\tilde{C}_l$ spectra which are not present in the average spectrum $<\tilde{C}_l> = \tilde{C}^l$. This procedure does not rely on guesswork, is straightforward and general, but the new $\{\tilde{C}_l\}$ are not very intuitive. The procedure bears some formal similarities with Gorski's construction [11], but the context and motivation are very different. Here we orthonormalize the $C_l$ basis with respect to the cosmic covariance matrix of the underlying theory (seen as an inner product). In fact our procedure can be applied to any starting basis, including Gorski's basis.

We will exemplify this procedure with textures. The covariance matrix for the $C_l$ in all the texture models considered in [4] can be found as a byproduct of the calculation in Sec.4 of [4]:

$$\mathrm{cov}(C_l, C_{l'}) = \frac{<a^2>}{4\pi}\left(1 + \frac{\sigma^2(a^2)}{<a^2>^2}\right)\int_0^{y_{ls}} dy\, N(y) W^{ls2}(y) W^{l's^2}(y) \qquad (10)$$

with $l \neq l'$. Insight into the correlations can be gained from the Pearson's correlation coefficient

$$\mathrm{cor}(C_l, C_{l'}) = \frac{\mathrm{cov}(C_l, C_{l'})}{\sigma(C_l)\sigma(C_{l'})} \qquad (11)$$

which always takes values in the range $[-1, 1]$, or the relative covariance matrix

$$\mathrm{rcov}(C_l, C_{l'}) = \frac{\mathrm{cov}(C_l, C_{l'})}{C^l C^{l'}} \qquad (12)$$

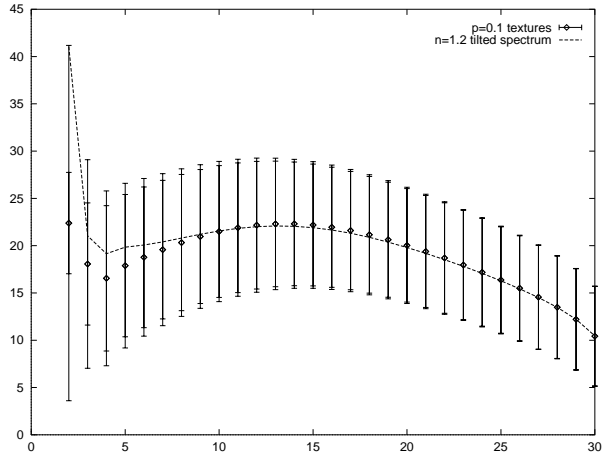

Figure 7: The $\tilde{C}_l$ spectrum for texture Gaussian spots with $p = 0.1$ and $n = 1$. The inner (outer) error bars are Gaussian (full) cosmic variance error bars. We have also shown the fitting $n^i = 1.2$ tilted theory $\tilde{C}_l$ spectrum.

which factors out the absolute size of the $C_l$. The cosmic covariance matrix for textures depends on $n$ and $p_s$, which should first be estimated from tilt and normalization measurements at intermediate $l$. In Fig. 5 we have plotted the off-diagonal inferior elements of the correlation matrix $\mathrm{cor}(C_l, C_{l'})$ for our favoured texture model. Note how the correlations decay away from the diagonal, and also the abnormally high correlations between the low $l$ multipoles. It is curious to note that correlations between neighbouring multipoles do not go to zero for high $l$ as fast as one might expect. All correlations become meaningless for large values of $n$ or $p_s$. Following [4] it can be proved that the covariance matrix conditional to $y_1$ is simply

$$\mathrm{cov}(C_l, C_{l'}|y_1) = \frac{<a^2>}{4\pi}\left(1 + \frac{\sigma^2(a^2)}{<a^2>^2}\right)\int_{y_1}^{y_{ls}} dy\, N(y) W^{ls2}(y) W^{l's^2}(y)\,. \qquad (13)$$

We have checked numerically that for all values of $y_1$ in the 68% confidence interval, the value of the Pearson's coefficient between adjoining low $l$ falls below 0.2 (as opposed to 0.8 for the unconditional $C_l$).

If $n$ and $p_s$ are large enough $\tilde{C}_l \approx C_l$, but as soon as the correlations become significant the following principal component structure emerges. Arranging the $\tilde{C}_l$ in decreasing order of cosmic variance, the first $\tilde{C}_l$ is a linear combination of the first few $C_l$ all multiplied by positive coefficients. This $\tilde{C}_l$ summarises the low $l$ normalization, and carries an abnormally large cosmic variance. The next few $\tilde{C}_l$ are approximately of the form $\frac{C_l - C_{l-1}}{\sqrt{2}}$, that is, they are derivative spectra ($\approx \partial C^l / \partial l$). As we go up in $l$ we recover the $\tilde{C}_l \approx C_l$ regime. The larger the $p_s$ the earlier we reach $\tilde{C}_l \approx C_l$. In Fig. 6, as an example, we have plotted the $\tilde{C}_l$ coordinates in $C_l$ space for a particular texture model ($n = 1$, $p_s = 0.1$ Gaussian spots), and one can see the principal component structure mentioned. In Fig. 7 we have plotted the corresponding $\{\tilde{C}^l\}$ spectrum with Gaussian and full cosmic variance error bars. No correlations exist for $\{\tilde{C}^l\}$, so comparison with experiment can now proceed in the usual way. There is still an excess of variance relative to Gaussian theories but, with the exception of the first $\tilde{C}^l$, the relative excess is now much smaller (see also Fig. 8). Again it seems that we are sweeping all the variance surplus under a single variable whenever we do away with the correlations.

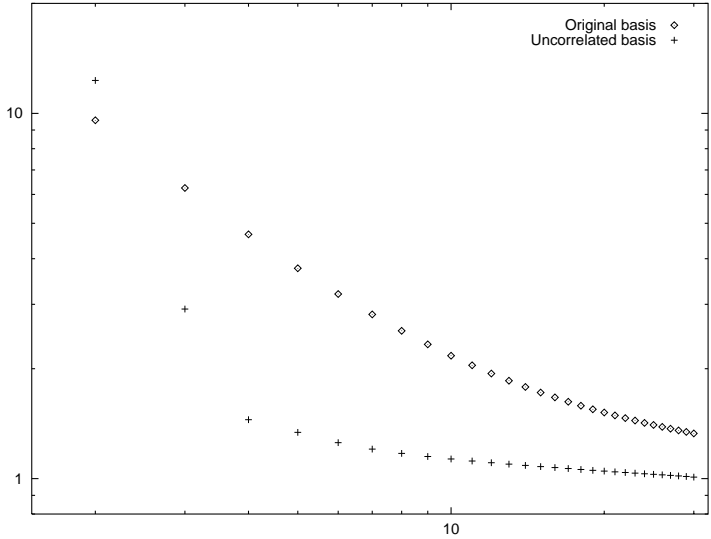

Figure 8: $\mathcal{V}_l$ for texture Gaussian spots ($p = 0.1$, $n = 1$) in the original $C_l$ frame and in the uncorrelated frame $\tilde{C}_l$.

## 3  Digression on the concept of predictivity

Besides their data analysis implications, the results in Section 2 also have something to say on the controversial issue of predictivity. Cosmic variance makes theories less predictive, but what this means mathematically is not altogether clear. Predictivity is a word colloquially used in serious verbal discussions which seldom makes its way into the literature. In Gaussian theories the $C_l$ are all independent and extensive (additive and positive). Hence it makes sense to define the variance of a spectrum as the sum of the $C_l$ relative variances

$$\mathcal{V} = \sum_{l=l_{min}}^{l_{max}} \frac{\sigma^2(C_l)}{C^{l2}} = \sum_{l=l_{min}}^{l_{max}} \frac{2}{2l+1} \ , \tag{14}$$

thus factoring out of the variance expression the absolute size of the spectrum. In non-Gaussian theories the $C_l$ are correlated variables and so this definition loses its meaning. Loosely speaking it is obvious that correlations reduce the spectrum variance from the sum of the $C_l$ marginal variances, but it is not easy to quantify this feeling. We may rotate the $C_l$ into an uncorrelated frame $\{\tilde{C}^l\}$, as in Sec. 2.2, but then the principal components are not extensive (for instance they may be negative). Therefore taking the relative variances is no longer a sensible way to factor out the absolute size of the spectrum. In fact not even in Gaussian theories does one have $\sigma^2(\tilde{C}_l) \propto \tilde{C}^{l2}$. A possible way out of this problem is to compare the variance in the $\tilde{C}_l$ with what this variance would be if the original $C_l$ were Gaussian random variables

$$\mathcal{V}_l = \frac{\sigma^2(\tilde{C}_l)}{\sigma_G^2(\tilde{C}_l)} \ . \tag{15}$$

From these quantities one can then define the spectrum variance relative to a Gaussian spectrum variance as

$$\mathcal{V}_r = \frac{1}{D} \sum_{l=l_{min}}^{l_{max}} \mathcal{V}_l \tag{16}$$

with $D = l_{max} - l_{min} + 1$. In this expression correlations have been duly taken into account and the spectrum size has been factored out. The predictivity relative to the Gaussian predictivity may then be defined as $\mathcal{P}_r = 1/\mathcal{V}_r$. We have computed these quantities for the texture $p_s = 0.1$ model and found $\mathcal{V}_r = 1.55$. Also in Fig. 8 we have plotted $\mathcal{V}_l$ for this model in the original $C_l$ frame and in the uncorrelated frame $\tilde{C}_l$. We see that we would have grossly overestimated the variance of the spectrum by neglecting the correlations. With the exception of the first (normalization) variable $\tilde{C}_l$, the $\mathcal{V}_l$ are much smaller in the new frame. It turns out that texture models are less predictive than Gaussian theories but not as much as overlooking correlations might have suggested. In general the lower the $p_s$ the less predictive textures are. However we have here considered only $C_l$ spectra. One must bear in mind that especially for low $p_s$ texture models the best predictions might be in the form of multipole shape factors or inter-$l$ correlators [9]. Gaussian theories, on the contrary, are maximally non predictive in this aspect.

# 4  Non-Gaussian signals in $\{C_l\}$ space

In Section 2 we tackled the problem of predicting $C_l$ spectra subject to cosmic covariance. Here we consider how to confront theories with possibly different $C_l$ covariance matrices. Clearly correlations cannot be neglected as they may constitute the main difference between one theory and another. It is generally impossible to find a single variable capturing most of the gap between the two theories. Therefore we devise comparison methods which make use of the joint distribution functions of the whole $\{C_l\}$ spectrum. Once more we explain our ideas by applying them to concrete examples. We confront the low-$l$ texture models we have used above with tilted spectra Gaussian theories which fit them at $l \in (25, 30)$. As shown in [4] this fit still leaves a significant suppression of power at low $l$ in texture models. We now quantify the strength of this signal.

It will be useful here to use the concept of cosmic confusion between two theories $T_1$ and $T_2$ in a set $Q$ of measurable quantities. This is a measure of the overlap of the distribution functions $F_1(Q)$ and $F_2(Q)$, and is essentially the percentage of the two populations which can be put in a one to one correspondence:

$$\mathcal{C}_Q(T_1, T_2) = \int dQ \ \min(F_1(Q), F_2(Q)) \ . \tag{17}$$

$\mathcal{C}_Q(T_1, T_2)$ varies between 0 (measuring $Q$ will act as a crucible between the two theories) and 1 ($T_1$ and $T_2$ are the same theory as far as $Q$ is concerned). As a guide to the meaning of $\mathcal{C}$ we can express it in terms of $n$-sigmas. This is defined as the separation in units of $\sigma$ between the peaks of two 1-D Gaussians with variance $\sigma$ which gives an overlap $\mathcal{C}$. More concretely $\mathcal{C} = \text{erfc}(n/(2\sqrt{2}))$. For a 1, 2, 3, 4 $\sigma$ differentiation one has 0.61, 0.31, 0.13, and 0.04 confusion, respectively. Given a set $Q$ of variables the confusion between two theories is invariant under non-singular transformations on $Q$. It can also be proved that if we ignore one of the variables in the set and marginalize the distributions $F_1$ and $F_2$ with respect to it we can only increase the cosmic confusion. Hence by considering the cosmic confusion between texture theories and their fitting tilted Gaussian theories in $C_l$ for all $2 \leq l \leq 30$ (a subset of all the variables) we obtain an upper bound on the confusion between the theories. By adding high-$l$ sections of the $C^l$ spectrum or $m$-structure spectra [9] into our predictions we can only decrease the confusion. Also a single low-$l$ variable may exist which captures most of the large angle gap between the theories, but the confusion in this variable will never be smaller than the confusion in the $C_l$ for all $2 \leq l \leq 30$.

Let us now look at $\{C_l\}$ spectra as points in a $D$-dimensional vector space (where $D = l_{max} - l_{min} + 1$). By varying the free parameters of texture and Gaussian models their average spectra $\{C^l\}$ span two 2-D surfaces in this space. One can set up a map between these two surfaces so as to maximize the cosmic confusion (a procedure approximated by the fit performed at $l \in (25, 30)$). Is the confusion left by this identification high enough to render low-$l$ spectra useless?

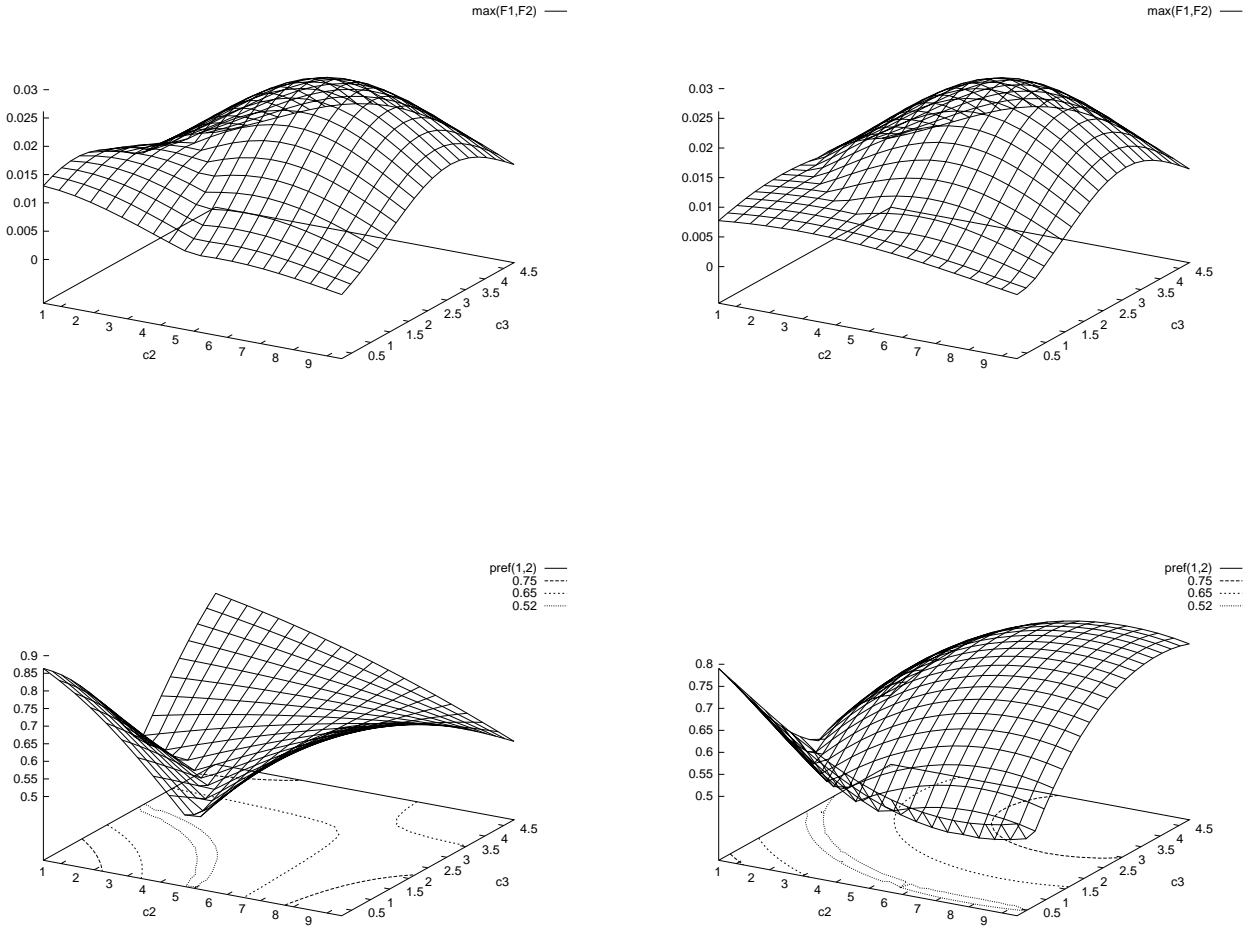

Figure 9: The functions $\max(F_{tx}(C_2, C_3), F_G(C_2, C_3))$ (top) and $\mathrm{Pref}(F_{tx}(C_2, C_3), F_G(C_2, C_3))$ (bottom) with (right) and without (left) texture correlations switched on. The base space lines are contours of iso-preference in the $C_2$-$C_3$ plane. Correlations are essential for bringing out the texture preference summit (compare the relative heights of the texture and inflation peaks on the left and on the right). Correlations also reconfigure the preference contours and turn the valley of indecision ($\mathrm{Pref} = 1/2$) into a gorge.

| $p_s$ | $n^i$ | $l=2$ | $l=2,3$ | $l=2,3,4$ | $l=2..7$ | $l=2..11$ |
|---|---|---|---|---|---|---|
| 0.05 | 1.3 | 0.42 | 0.33 | 0.25 | 0.15 | 0.114 |
| 0.1 | 1.2 | 0.70 | 0.58 | 0.52 | 0.39 | 0.32 |
| 0.15 | 1.12 | 0.71 | 0.60 | 0.55 | 0.46 | 0.41 |
| 0.25 | 1 | 0.60 | 0.48 | 0.42 | 0.355 | 0.327 |

Table 1: Cosmic confusion in various sets of variables between four texture theories and their fitting tilted spectra Gaussian theories (fit performed at $l \in (25, 30)$).

Plots like Fig. 1 are misleading as they ignore $C_l$ correlations and suggest that cosmic variance error bars are hypercubes in the $\{C_l\}$ space. In fact $C_l$ error bars are always hyperovaloids, $D-1$ dimensional surfaces of equal probability inside which a given percentage of the population lives. The principal axes of the ovaloid are parallel to the cartesian axes only when the $C_l$ are uncorrelated, but even then the ovaloid axes dimensions are not the marginal variances as plotted in Fig. 1. We are therefore dealing with a $D$-dimensional problem which can never be factorized into $D$ one-dimensional problems. Due to the complexity of the problem we decide here to truncate the analysis at the level of the second moments of the $C_l$ distribution. In this approximation the $C_l$ distribution is approximated by a multivariate Gaussian distribution with the theory's covariance matrix. The ovaloidal error bars become ellipsoidal and one must extend the $C_l$ range to $C_l \in (-\infty, \infty)$. This is a rough approximation, not true even for Gaussian theories, for which the joint $C_l$ distribution is a product of $\chi_{2l+1}$ 1-D distributions. However, this approximation does put texture and Gaussian theories at the same level of approximation while allowing for texture non-Gaussian features to be included, in the form of off-diagonal elements in the covariance matrix. In Table 1 we show the cosmic confusion between various texture models and their fitting tilted Gaussian models in various sets of $C_l$ for $2 \leq l \leq 11$, computed in this approximation. In all cases the confusion in $C_l$ for $11 \leq l \leq 30$ is nearly 1. Although the largest single $C_l$ differentiation between the two theories occurs for $C_2$, one has to take the first ten $C_l$ into account to achieve a 2-sigma signal.

Once a measurement is made the situation changes. Imagine first that one makes a measurement of the set $Q$ with infinite experimental accuracy. If $F_1(Q) > F_2(Q)$ we can then say that the probability of theory $T_1$ over theory $T_2$ is

$$\text{Pref}_{12}(Q) = \frac{\max(F_1, F_2)}{F_1 + F_2}, \tag{18}$$

the preference function. The confusion after the measurement is now

$$\mathcal{C} = \frac{\min(F_1, F_2)}{2(F_1 + F_2)} = 2(1 - \text{Pref}_{12}). \tag{19}$$

The preference function foliates the $\{C_l\}$ space into preference contours. $\text{Pref}_{12} = 1/2$ represents the valley of indecision. Summits represent the most conclusive possible results. There should be at least two summits, one for each theory. Imagine now that a measurement is made giving an intermediate $l$ best fit at $n^i = 1.2$ for inflation, $p_s = 0.1$ for textures. Let us examine how the low $C_l$ could act as a referee. For graphical purposes we have confined ourselves to the $C_2 - C_3$ plane, and in Fig. 9 we have plotted the preference function, and its iso-contours. Notice how different these are if one switches off the correlations. It turns out that correlations are essential for pulling the texture summit up to the same height as the inflationary summit. Also the valley of indecision has much steeper surrounding slopes if texture correlations are taken into account. Since the prominent signal associated with textures is only there because of the correlations, this signal is to be seen as a sign of non-Gaussianity.

Naturally, the outcome of the experiment is a domain of results with an experimental distribution function. Therefore given a particular experiment, the probability of one theory over another (or the experimental cosmic confusion) is obtained by integrating expression (18) weighted by the experimental distribution over the whole domain.

## 5 The message

This paper is intended as a warning. By custom and tradition the average $C^l$ spectrum is always taken as a sensible prediction for what the observed $C_l$ should be. We have shown with examples how non-Gaussian statistics may render the $C^l$ a naive and misleading prediction for the observed sky. If you do not have any reason to postulate Gaussianity then we recommend the following recipe. Start by computing the $C^l$ and then always compute the $C_l$ variance. Compare it with the Gaussian variance. If a large excess variance exists, then the standard recipe does not work. You are dealing with a cosmic covariance problem, and in this paper we gave some alternative recipes for making predictions and confronting theories. A topical example is defect Doppler peaks [12]. Suppose that an excess variance is found for the relevant $C_l$. Then, even if the average $C^l$ shows a single bump, this might have little to do with what any observer sees. It could happen that each observer sees a rich peak structure, similar to the inflation peaks, but the peaks' position and height could be random variables (in fact, random indices). That being the case, the average $C^l$ would be the statistically weighted envelope of all the possible peak curves, and the $C^l$ would be blind to the peak structure seen by any observer. Computer simulations performed conditional to a value of the random indices would show a system of peaks. A calculation averaging over the whole ensemble would not. Whatever one's methods, in such a scenario the sensible prediction would be the $C^l$ spectrum inside each sub-ensemble of fixed random indices, the residual cosmic variance, and the random indices distribution.

The author feels that this type of behaviour (transposed to the context of sample variance) might be the entire point in the controversy surrounding the apparently contradictory results given by small angle detections [13].

# Acknowledgements

I would like to thank P.Ferreira and M.P.Hobson for discussion and helpful suggestions, and K.Baskerville for help in the preparation of the manuscript. This work was supported by a research fellowship at St.John's College, Cambridge.